\documentclass[conference,letterpaper]{IEEEtran}

\usepackage[utf8]{inputenc}
\usepackage[T1]{fontenc}
\usepackage{graphicx}
\usepackage[caption=false,font=footnotesize]{subfig}
\usepackage[cmex10]{amsmath}  % [cmex10] for IEEE Xplore compliance
\usepackage{amsfonts}
\usepackage[dvipsnames]{xcolor}
\usepackage{url}
\usepackage{cite}
\usepackage{ifthen}
\usepackage{bm}

\usepackage[ruled]{algorithm}
\usepackage{algpseudocode}
\usepackage{comment}

\usepackage{ioa_code}

\algrenewcommand\algorithmiccomment[1]{\hspace{2em}{\color{olive} // \textit{#1}}}

\newcommand{\Nat}{\ifmmode \mathbb{N} \else $\mathbb{N}$ \fi}

\newcommand{\Real}{\ifmmode \mathbb{R} \else $\mathbb{R}$ \fi}

\newcommand{\Fin}[1]{\ifmmode \mathbb{F}_#1 \else $\mathbb{F}_#1$ \fi}

\newcommand{\tup}[1]{%
    \relax\ifmmode
      \langle #1 \rangle%
    \else
        $\langle$#1$\rangle$%
    \fi
}

\newcommand{\act}[1]{%
    \relax\ifmmode
        \mathord{\mathcode`\-="702D\sf #1\mathcode`\-="2200}%
    \else
        $\mathord{\mathcode`\-="702D\sf #1\mathcode`\-="2200}$%
    \fi
}

\newcommand{\remove}[1]{}

\makeatletter
\def\mainlistofsymbols{
  \normalsize
  \vspace*{1.5 em}
  \@starttoc{los}
}

\def\partonelistofsymbols{
  \normalsize
  \vspace*{1.5 em}
  \@starttoc{p1los}
}

\def\parttwolistofsymbols{
  \normalsize
  \vspace*{1.5 em}
  \@starttoc{p2los}
}

\def\l@symbol#1#2{\addpenalty{-\@highpenalty} \vskip 4pt plus 2pt
{\@dottedtocline{0}{0em}{8em}{#1}{#2}}}
\makeatother

\newcommand{\newhiddensym}[2]{%
}

\newcommand{\algIOA}[2]{\ifmmode{\text{#1}_{#2}}\else{$\text{#1}_{#2}$}\fi}
\newcommand{\EX}{\ifmmode{\xi}\else{$\xi$}\fi}
\newcommand{\EXF}{\ifmmode{\phi}\else{$\phi$}\fi}
\newcommand{\inter}[1]{
	\ifmmode{\left(\bigcap_{\mathcal{Q}\in#1}\mathcal{Q}\right)}
	\else{$\left(\bigcap_{\mathcal{Q}\in#1}\mathcal{Q}\right)$}
	\fi
}
\mathchardef\mhyphen="2D
\newcommand{\vid}[1]{\ifmmode{\nu_{#1}}\else{$\nu_{#1}$}\fi}
\newcommand{\seen}{\ifmmode{seen}\else{$seen$}\fi}
\newcommand{\maxts}[1]{\ifmmode{maxTS_{#1}}\else{$maxTS_{#1}$}\fi}
\newcommand{\maxtag}[1]{\ifmmode{maxTag_{#1}}\else{$maxTag_{#1}$}\fi}
\newcommand{\maxpair}[1]{\ifmmode{maxMPair_{#1}}\else{$maxMPair_{#1}$}\fi}
\newcommand{\mintag}[1]{\ifmmode{minTag_{#1}}\else{$minTag_{#1}$}\fi}
\newcommand{\maxps}{\ifmmode{maxPS}\else{$maxPS$}\fi}
\newcommand{\conftg}[1]{\ifmmode{confirmed_{#1}}\else{$confirmed_{#1}$}\fi}
\newcommand{\maxconftag}{\ifmmode{\ms{maxCT}}\else{$maxCT$}\fi}

\algblockdefx[Operation]{Operation}{EndOperation}%
[2]{{\bf operation} $\act{#1}$(#2)}%
{{\bf end operation}}
\algblockdefx[Procedure]{Procedure}{EndProcedure}%
[2]{{\bf procedure} $\act{#1}$(#2)}%
{{\bf end procedure}}
\algblockdefx[Receive]{Receive}{EndReceive}%
[2]{{\bf Upon receive} (#1)$_{\text{ #2 }}${\bf from} $q$}%
{{\bf end receive}}

\title{Optimum Peer-Turbo: A Scalable and Efficient Solution for P2P Broadcasting \vspace{-.5 cm}}

\author{
Muriel M\'edard\textsuperscript{1}
Kishori M. Konwar\textsuperscript{1},
Moritz Grundei \textsuperscript{1},
Vipindev Adat Vasudevan\textsuperscript{2}\\
\IEEEauthorblockA{\textsuperscript{1}\texttt{\{mmedard, kkonwar, moritz\}@getoptimum.xyz}}
\IEEEauthorblockA{\textsuperscript{1}Optimum, Cambridge, MA, USA}
\IEEEauthorblockA{\textsuperscript{2}\texttt{vipindev@mit.edu}}
\IEEEauthorblockA{\textsuperscript{2}Massachusetts Institute of Technology, Cambridge, MA, USA}
\vspace{-1 cm}
}

\begin{document}

\maketitle

%%% If eligible for the student paper award, add this as the first line
%%% of the abstract (remove from final accepted version):
%%% THIS PAPER IS ELIGIBLE FOR THE STUDENT PAPER AWARD.

\begin{abstract}

Blockchain systems such as Solana or Monad employ tree- or star-shaped broadcast topologies in which a single source node disseminates message shards to a set of target peers within a strictly bounded time window. In these architectures, shard propagation must complete before the next consensus step, making timely delivery to a large fraction of the validator set essential. A fundamental limitation of such designs is that the outbound bandwidth of the source node constitutes the primary system bottleneck. In this paper, we introduce peer Turbo, a technique that allows target nodes to exchange shards using Random Linear Network Coding (RLNC), thereby assisting each other in completing decoding without requiring explicit shard state coordination. We use a tractable fluid approximation of the degree of freedom distribution of peer-Turbo-enabled systems show that this approach reduces source bandwidth required for a set service quality by up to one order of magnitude, or equivalently reduces propagation latency by one order of magnitude under fixed bandwidth constraints.
\end{abstract}

\begin{IEEEkeywords}
  P2P Broadcast; RLNC; Erasure Coding;
  %\vspace{-.5 cm}
\end{IEEEkeywords}

\section{Introduction}
\label{sec:intro}
%Consider systems where a sender broadcasts in a hierarchical fashion to destination nodes with the goal of low latency to all or some fraction of the  destination nodes. Such propagation occurs from proposer to attesters in Solana Rotor or Turbine, or  in the Raptorcast system of Monad. That model also represents well the BNB
%propagation for the execution layer.

%We have thus a framework where
%a single sender transmits along a hierarchical system to a set
%of receivers. In the simplest case, such as Rotor, the system is a star topology. Notably absent is the assistance in a peer-to-peer 
%approach of the destination nodes. 
%Owing to the use of
%structured codes such as RS, or of no coding at all, the absence
%of mutual assistance among nodes may be due to the fact that
%such cooperation would require state information of the shards
%obtained. Without such state information, duplicate shards
%are highly likely in such systems, since all of the receivers
%see shards being sent in the same order from the sender. The
%sharing of state information is onerous and leads to delays that
%run against the low latency that is sought in the systems we
%consider.

%In this work we propose a different approach, based
%on the fact that when using RLNC, even when receivers
%experience a significant overlap in the received shards, nodes
%can assist each other without the need for state information
%mentioned above \cite{Lucetal09}.

% Vipindev - This introduction looks good but is lengthy. We probably have nothing to cut off technically, but just to shorten it. 

\begin{figure*}[!t]
    \centering
    %\vspace{-2 cm}
    \subfloat[Without Peer-Turbo: a single source disseminates shards directly to $m$ target peers.\label{fig:basic}]{%
        \includegraphics[width=0.45\linewidth]{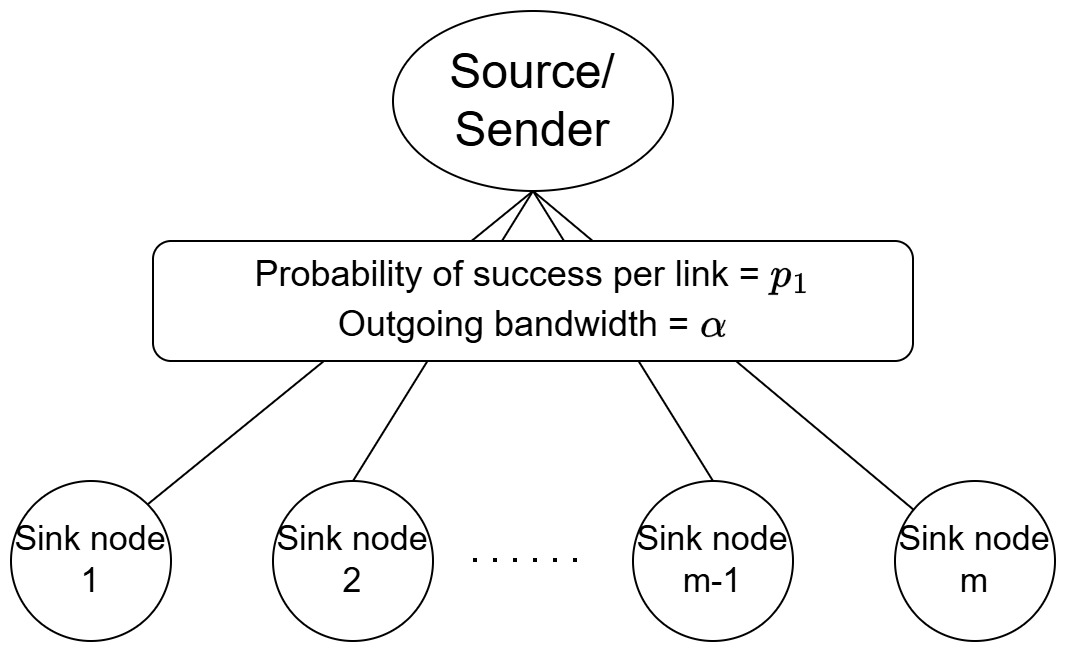}%
    }
    \hfill
    \subfloat[With Peer-Turbo: Peers supplement shards disseminated by the source node with RLNC coded shards shared among each other.\label{fig:peer_turbo}]{%
        \includegraphics[width=0.45\linewidth]{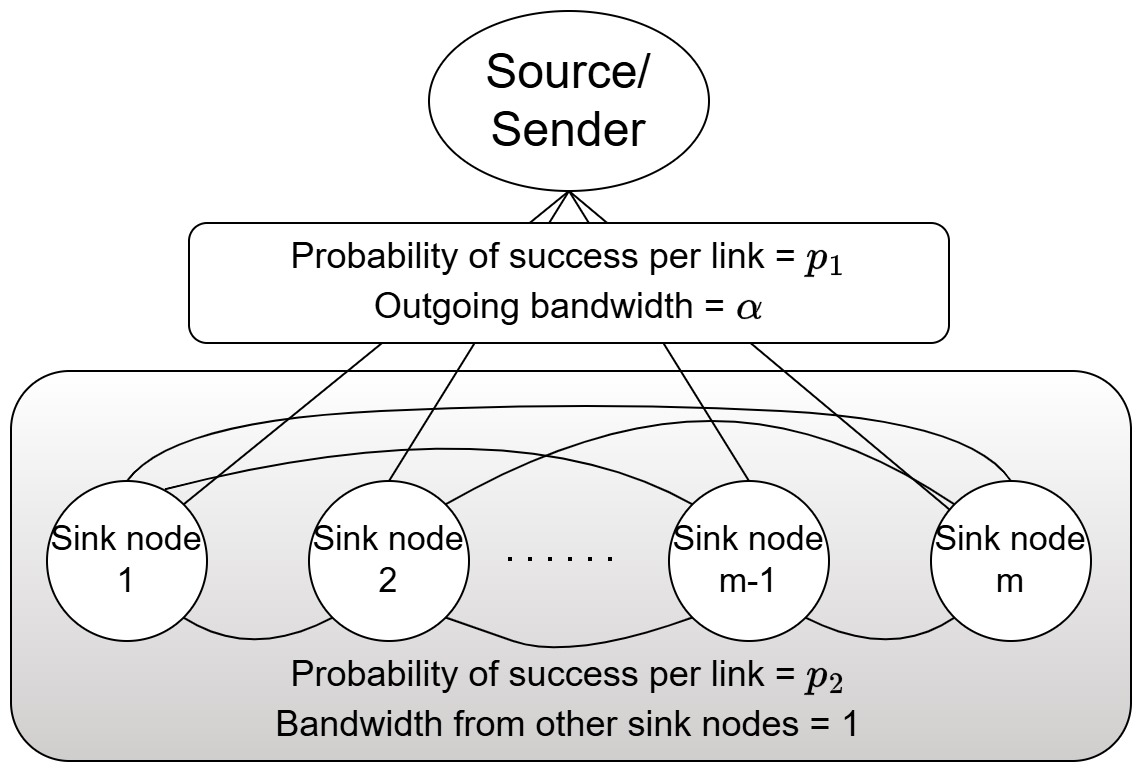}%
    }
    \caption{System Model. \ref{fig:basic} shows the current systems without Peer-turbo while \ref{fig:peer_turbo} shows the proposed peer-turbo approach.}
    \vspace{-.5 cm}
    \label{fig:system_models}
\end{figure*}

Blockchain systems are distributed state machines that maintain shared state across mutually untrusted nodes. Consensus protocols coordinate state transitions through message exchange, typically including dissemination of a proposed payload from a source, or proposer, to target peers such as validators, as in Solana, Ethereum, and Monad~\cite{solana,monad,ethereum2}. Current systems mainly use two dissemination paradigms. The first is mesh-based gossip, where each node maintains a bounded, irregular peer set; Ethereum's Gossipsub is a prominent example~\cite{gossipsub-specification}. Although gossip reduces per-node bandwidth requirements, payloads must traverse multiple hops, which can lower effective throughput in uncoded, unsharded systems as packet loss compounds and latency increases. Recent proposals therefore introduce coding into Ethereum's networking layer to add controlled redundancy and improve resilience to packet loss. In particular, ethp2p, scheduled to succeed Gossipsub in late 2026, proposes Reed--Solomon erasure coding for disseminating large Ethereum objects under partial packet loss~\cite{ethp2p-reed-solomon,reed1960polynomial}.

The second paradigm, which we focus on in this paper, uses tree or star topologies where the source disseminates data either directly to target peers or hierarchically through relays, often after splitting the payload into shards. Examples include Solana's Turbine and Rotor protocols~\cite{solana,alpenglow_whitepaper}: in Rotor, a leader Reed--Solomon (RS) encodes a block into shards and distributes them to relay subsets, which forward them to roughly $1300$ validator target peers~\cite{alpenglow_whitepaper}. Similar designs appear in Monad's RaptorCast, which combines hierarchical dissemination with rateless Raptor codes~\cite{monad,shokrollahi2006raptor}, and in high-performance execution-layer networks such as BNB Chain, where proposers disseminate large execution payloads under strict timing constraints~\cite{bnb_whitepaper}. Although these architectures reduce hop count and latency relative to mesh gossip, they remain bottlenecked by the source node's outbound bandwidth.

In this work, we propose a modular improvement to star-based dissemination architectures that preserves their low-latency properties while reducing source bandwidth requirements independently of the coding scheme used by the underlying p2p broadcast protocol. We introduce peer Turbo, which allows destination nodes to exchange information during dissemination, as shown in Fig.~\ref{fig:peer_turbo}, to accelerate decoding. To avoid excessive coordination overhead, peer Turbo uses random linear network coding (RLNC)~\cite{RLNC2006,li2003linear}, which has been widely studied for distributed storage~\cite{abdrashitov2015durable,deb2005good}, hybrid storage-code systems~\cite{babelstorage,mulitcodestorage}, and more recently blockchain dissemination and storage~\cite{ethresearch-rlnc-artice,nicolaou2025optimump2p,pricing}. 
%Unlike structured erasure codes such as RS codes, RLNC lets peers generate innovative information without knowing which shards others already hold, eliminating shard-state synchronization and the associated signaling delays that could otherwise undermine star-based low-latency dissemination.

The core contribution of this paper is a tractable fluid approximation over degrees of freedom, which we use to derive network level decoding latencies. This framework enables us to characterize the time required for a given fraction of nodes to collect enough innovative shards to decode the payload. We further underline the advantages of peer Turbo by providing numerical results that show up to 10x reductions in decoding latency under the same source outbound restrictions.

% \section{Notation}
% \label{sec:notation}
% \input{sec-notation}

\section{System Model}
\label{sec:model}
\begin{figure*}[htbp]
    \centering
    %\vspace{-2 cm}
    \includegraphics[width=\linewidth]{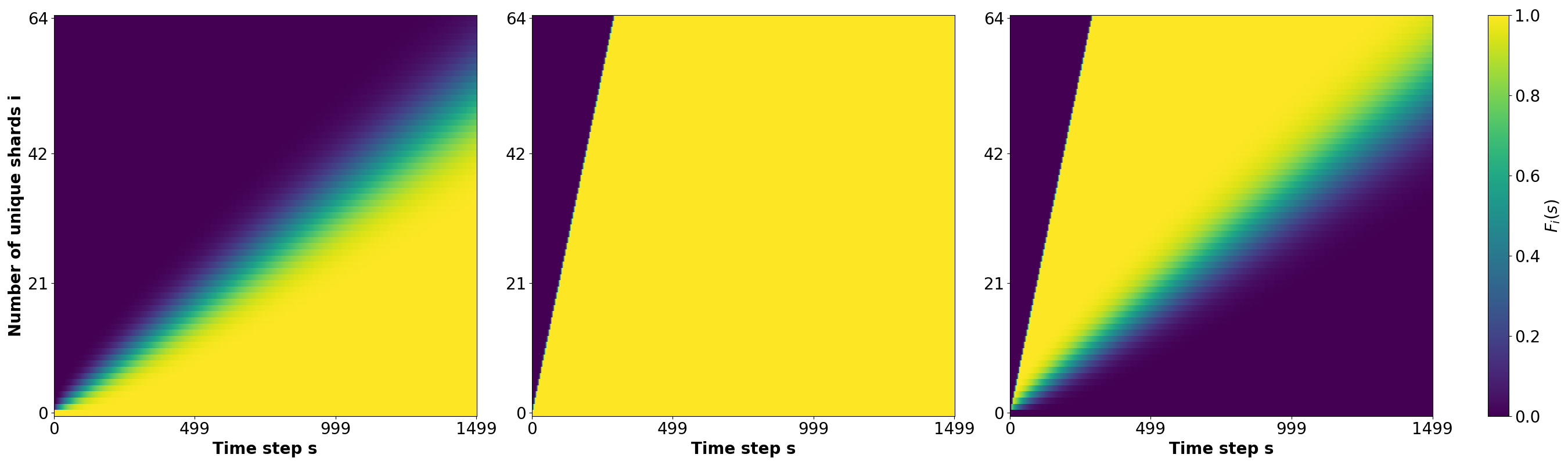}
    \caption{Empirical survival function (SF) for number of unique shards after a number of time steps $s$ without peer-Turbo connections (left), with peer-Turbo connections (center) and the corresponding difference in survival functions (right) according to eqs.~\eqref{eqn1} and \eqref{eq:peer_turbo_recurrence}. For both simulations, link reliabilities of $p_{1/2}=0.9$, $\alpha=50$ and $m=1300$ was assumed.}
    \vspace{-.5 cm}
    \label{fig:heatmap}
\end{figure*}

We consider a dissemination model in which a single node, called the source, transmits a payload to $m$ target nodes. The payload is divided into $k$ data shards $\bm{x}_1,\dots,\bm{x}_k\in\mathbb{F}_q^l$, and the source disseminates coded shards $\bm{c}_j\in\mathbb{F}_q^l$ to the target peers. For the analysis, we consider a single-layered, star-shaped topology in which the source has a direct connection to each target node. We remain agnostic to the exact code rate and treat the coded shard stream as effectively rateless, since the reception of any $k$ suitable shards is sufficient to decode the payload.

The outgoing bandwidth of the source is limited. We denote this total outgoing bandwidth by $\alpha$, and assume that it must be shared among the $m$ target nodes. Thus, as the number of targets grows, the source cannot allocate a dedicated bandwidth resource to each target independently. Each transmission from the source to a target node is successful with probability $p_1$, and we model transmissions over different source-target links as independent Bernoulli trials. We denote by $p_2$ the success probability of a transmission over a link between two target nodes.

We model a discrete-time system with rounds. In each round, the source attempts to send coded shards to the target nodes through its available outgoing bandwidth. 
We denote by $\mathcal{S}(\bm{a}_1,\dots,\bm{a}_n)$ the subspace of $\mathbb{F}_q^\ell$ spanned by the vectors $\bm{a}_1,\dots,\bm{a}_n \in \mathbb{F}_q^\ell$. We denote the dimension of
this subspace by $\dim \mathcal{S}(\bm{a}_1,\dots,\bm{a}_n)$. 
In the peer-Turbo setting, target nodes additionally exchange coded shards with one another. By time $t$, we assume that $s$ rounds have been completed and that $s \times \Delta t$ seconds have elapsed, where $\Delta t$ is the duration of one round. At any time step, we denote the subspace of received shards at target peer $j$ as $\mathcal{S}_j(s)$.

The central modelling idea is that we do not track individual shard transmissions or the exact evolution of $\mathcal{S}_j(s)$ for all $m$ target peers. Instead, we track the number of degrees of freedom available at each target node $j$ as $|\mathcal{S}_j|$. We then approximate the population-wide evolution of these subspace dimensions through a fluid approximation. This perspective is inspired by Kleinrock's classical M/G/1 analysis \cite{kleinrock1975queuing}: rather than following the discrete trajectory of each individual object, we derive recursive equations for the aggregate evolution of the relevant state variable. The degree-of-freedom fluid approximation is not specific to payload broadcast in star-shaped topologies. It is also well suited to coded storage and content-delivery settings, such as \cite{babelstorage}, where content is retrieved from multiple coded storage systems with minimal coordination overhead. In such systems, the central state variable is not the identity of individual shards, but the amount of innovative coded information available to a receiver. A fluid approximation over degrees of freedom can therefore provide a tractable way to study population-level decoding progress, latency, and transmission overhead without explicitly tracking all fragment types, identities or coordination decisions.

To study population-wide shard state distribution, we introduce an empirical survival function over the target nodes. Let 
\begin{equation}
    F_i(s) \equiv \frac{1}{m}\sum_{j=1}^m\mathbb{I}(|S|_j\geq i)
\end{equation} 
.Thus, $F_0(0)=1$ and $F_i(0)=0$ for all $i>0$. The fraction of target nodes whose received subspace has exactly dimension $i$ after round $s$ is given by $G_i(s) = F_i(s) - F_{i+1}(s)$.

This representation allows us to study the population-wide evolution of decoding progress through the distribution of degrees of freedom across target nodes. We now derive discrete-time approximations for this evolution in two settings: first without peer-Turbo, where only the source transmits to the target nodes, and then with peer-Turbo, where target nodes can also assist one another.

\subsection{Without Peer-Turbo}

When there is no peer-Turbo, any destination node only receives new shards directly from the source in a round, as in Fig.~\ref{fig:basic}. The probability of a node getting a shard is determined by the success probability $p_1$ and the capacity of each link $\alpha/m$. Thus, we have the following evolution for $F$:

\begin{equation}\label{eqn1}
    \small F_i(s+1)= \min \left(  p_1\frac{\alpha}{m}\left( F_{i-1}(s) - F_i(s) \right) + F_i(s), 1 \right)
\end{equation}

where the first term in the RHS represents the fraction of nodes that raise their subspace dimension from $i-1$ to $i$ thanks to the reception of equations from the source and the second term represents the fact that nodes that already have a subspace of dimension at least $i$ continue to do so at $s+1$. The maximum value of $F_i(s)$ is 1.

For a system without peer-Turbo, we can define the fraction of nodes with at least $i$ dimensions in terms of the link success probability and the fraction of nodes with exactly $i$ subspace dimensions after a given round $s$. The derivation is as follows:
\begin{align*}
    G_i(s) &= F_{i-1}(s) - F_i(s) \\
    F_i(s+1)-F_i(s) &= p_1\frac{\alpha}{m} G_i(s) \\
    F_i(s) &= {p_1}\frac{\alpha}{m} \sum_{s=1}^t G_i(s-1)
\end{align*}

\begin{figure*}[!t]
    \centering
    %\vspace{-2 cm}
    \subfloat[Impact of source bandwidth multiple $\alpha$ ($k=32$, $p=0.9$, $m=1300$).\label{fig:sf_var_alpha}]{%
        \includegraphics[width=0.48\linewidth]{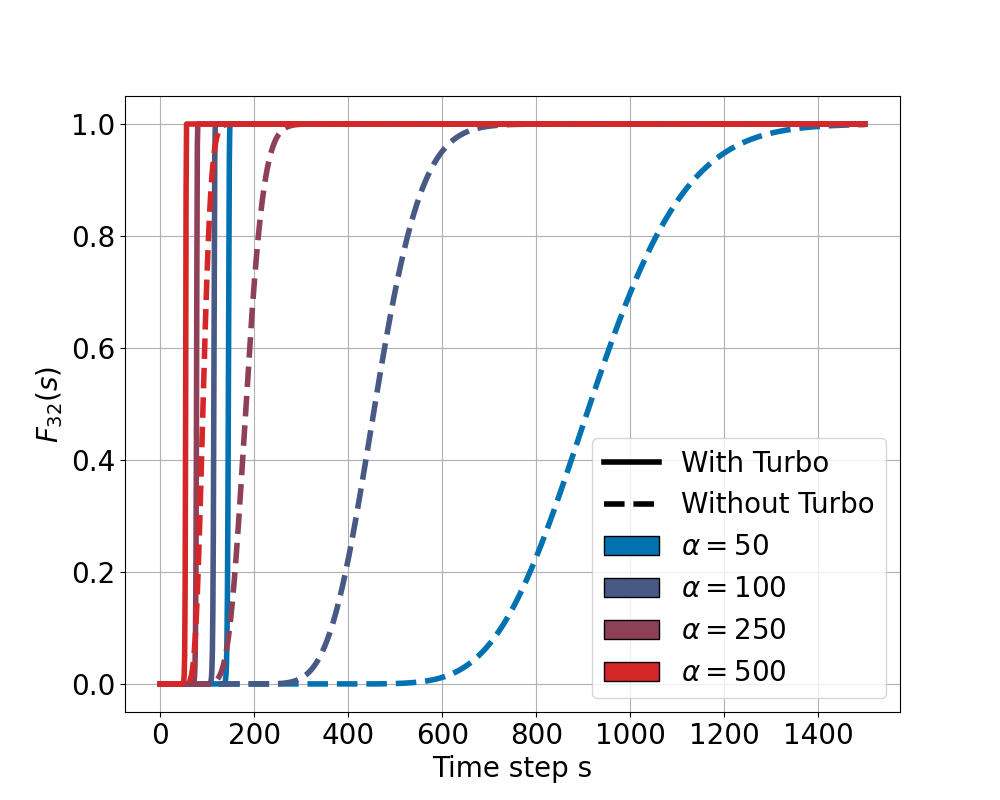}%
    }
    \hfill
    \subfloat[Impact of number of peers $m$ ($k=32$, $p=0.9$, $\alpha=50$).\label{fig:sf_var_m}]{%
        \includegraphics[width=0.48\linewidth]{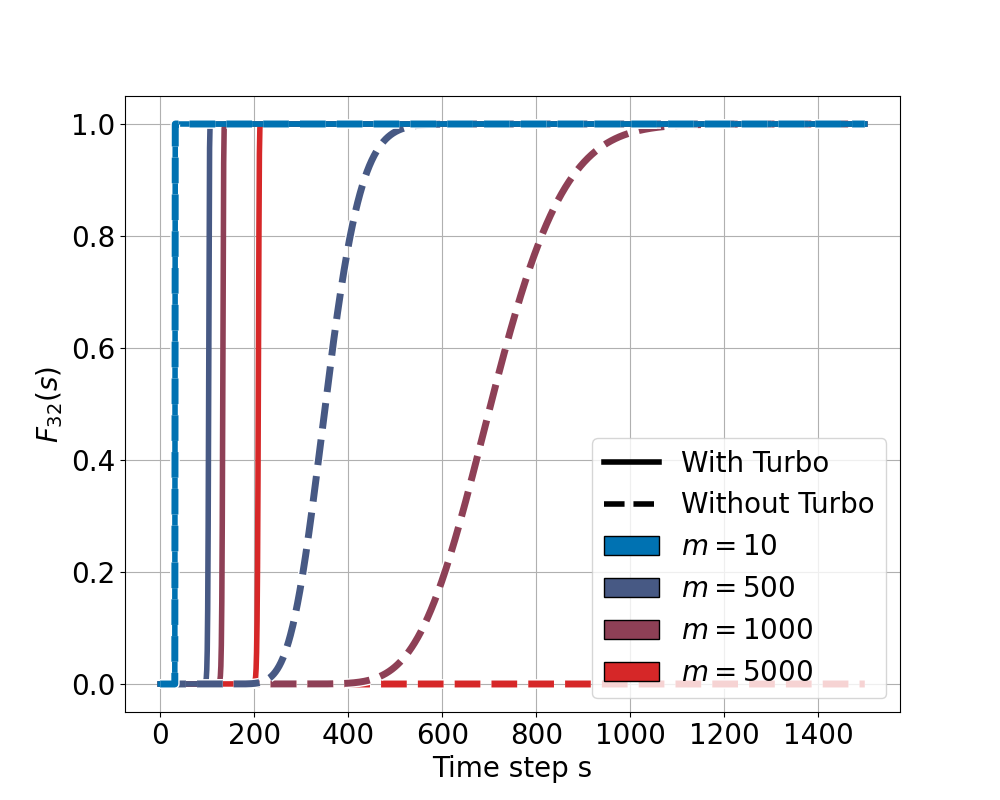}%
    }
    \caption{Empirical survival functions for setups without and with peer-Turbo connections under varying system parameters.}
    \vspace{-.5 cm}
    \label{fig:sf_sensitivity}
\end{figure*}

\subsection{With Peer-Turbo}

Peer-Turbo enables target nodes to cooperate by exchanging coded shards among themselves, as illustrated in Fig.~\ref{fig:peer_turbo}. We use a conservative model for peer assistance: a peer $j$ is assumed to provide useful information to a target node $i$ only if its received subspace has strictly larger dimension, i.e., $|S_j(s)| > |S_i(s)|$.
This gives a lower-bound estimate of the benefit of peer cooperation. In practice, even peers with subspaces of equal dimension may still be mutually useful whenever their subspaces are not identical.

To reduce the probability of non-innovative transmissions, target peers use random linear network coding (RLNC). Specifically, peer $j$ generates a recoded shard as
\begin{align*}
c_{\mathrm{RLNC}}^{(j)} = \sum_{\ell=1}^{|\mathcal{S}_j(s)|} \alpha_\ell c_\ell^{(j)},
\end{align*}
where $\mathcal{S}_j(s)=\{c_1^{(j)},\ldots,c_{|\mathcal{S}_j(s)|}^{(j)}\}$ denotes the set of innovative shards currently held by peer $j$, and the coefficients $\alpha_\ell \in \mathbb{F}_q$ are chosen uniformly at random. Importantly, the shards being recoded need not themselves have been generated by RLNC: they may originate from structured erasure codes such as RS codes or from previous random recodings.

Let $\epsilon_{i \leftarrow j,s}$ denote the probability that a successfully received recoded shard from peer $j$ lies in the existing span $S_i(s)$ of target node $i$ and therefore fails to provide a new degree of freedom. If $\dim S_j(s)>\dim S_i(s)$, then a random linear combination drawn from $S_j(s)$ falls inside $S_i(s)$ with probability at most
\begin{align*}
    \epsilon_{i \leftarrow j,s} \leq q^{-(\dim S_j(s)-\dim S_i(s))} \leq \frac{1}{q}.
\end{align*}
Thus, for sufficiently large field size, the probability of receiving a non-innovative shard from a peer with more degrees of freedom becomes negligible. This is consistent with standard RLNC analyses, where decoding failure and rank-deficiency probabilities vanish as the finite-field size increases~\cite{robust_network_codes,decoding_failure_prob,li2017hierarchical}. We therefore approximate each successful transmission from a peer holding more degrees of freedom as contributing one additional degree of freedom. Under this approximation, RLNC-recoded shards are especially useful for peer assistance: instead of forwarding only fixed coded shards, a peer can continuously generate fresh combinations that are innovative with high probability and hence likely to contribute directly to decoding~\cite{pricing}.

Consider now the evolution of $F_i(s)$, the fraction of target nodes with at least $i$ degrees of freedom after round $s$. Nodes that already belong to $F_i(s)$ remain in this set. The nodes that can newly enter $F_i(s)$ are those with exactly $i-1$ degrees of freedom. We denote this fraction by $G_i(s)=F_{i-1}(s)-F_i(s)$.

Such nodes can obtain a new degree of freedom either directly from the source or from a peer with a higher-dimensional subspace. The source contribution is proportional to
\begin{align*}
    %\vspace{-.2 cm}
    \frac{p_1\alpha}{m}G_i(s),
\end{align*}
while the peer-assisted contribution is proportional to
\begin{align*}    
    %\vspace{-.2 cm}
    p_2F_i(s)G_i(s),
\end{align*}

because, under the conservative assumption above, the fraction of potentially helpful peers is at least $F_i(s)$. Hence, for $1\leq i\leq k$, the degree-of-freedom fluid approximation is
\begin{equation}
%\begin{aligned}
    % F_i(s+1)
    % =
    % \min\Bigg(
    %     &F_i(s)
    %     + {}\\
    %     &\left(
    %         \frac{p_1\alpha}{m}
    %         +
    %         p_2F_i(s)
    %     \right)
    %     G_i(s),
    %     1
    % \Bigg).
    \small F_i(s+1) = \min\Bigg(F_i(s) + \left(\frac{p_1\alpha}{m} + p_2F_i(s) \right) G_i(s), 1 \Bigg).
%\end{aligned}
\label{eq:peer_turbo_recurrence}
\end{equation}

Equivalently,
\begin{align*}
    F_i(s+1)-F_i(s)
    &=
    \left(
        \frac{p_1\alpha}{m}
        +
        p_2F_i(s)
    \right)
    G_i(s),
    \\
    G_i(s)
    &=
    F_{i-1}(s)-F_i(s).
\end{align*}

The first term captures direct source transmissions, while the second term captures the additional degrees of freedom obtained through peers. Since the recurrence only counts peers with strictly larger subspace dimension as useful, it provides a conservative estimate of the benefit of Peer-Turbo.

%Let $\phi$ denote the target fraction of nodes that must decode %the payload. The corresponding dissemination latency is
%\begin{equation}
%    \tau
%    =
%    \operatorname*{arg\,min}
%    \left\{
%        t : F_k(t) \geq \phi
%    \right\}.
%    \vspace{-.2 cm}
%\end{equation}

\section{Numerical Results}
\label{sec:results}
We now evaluate the impact of peer-Turbo on plain message broadcast in systems with star topologies. Overall, peer-Turbo systems exhibit reduced tail latencies and enable significant reductions in the required source bandwidth $\alpha$ compared to systems that do not employ peer-Turbo connections.

First, we present the numerical evaluation of the peer-Turbo setup described in eq.~\eqref{eq:peer_turbo_recurrence} and compare it against systems that do not employ RLNC-enabled Turbo connections, as described in eq.~\eqref{eqn1}, using a mean-field simulation.

Fig.~\ref{fig:heatmap} displays the empirical survival function $F_i(s)$ for setups without a Turbo connection (left) and for setups in which sink nodes have a Turbo connection (center). The rightmost heatmap shows the difference between the two survival functions.

We observe that for both setups the empirical survival function is monotonous in time for any value of unique shards $i$, since the number of unique shards at any node can only increase during the dispersal phase. In systems without peer-Turbo connections, the transition between the regime where no peer holds $i$ shards and the regime where all peers hold $i$ shards becomes increasingly diffuse as $i$ grows while the peer-Turbo system exhibits a sharply defined transition even for larger values of $i$. The widening gap in the difference heatmap for increasing shard numbers reflects the growing benefit of peer-Turbo as losses compound over time.

Figs.~\ref{fig:sf_var_alpha} and~\ref{fig:sf_var_m} show that peer-Turbo substantially reduces sensitivity to both source bandwidth and network size. When the bandwidth multiple decreases from $\alpha=500$ to $\alpha=50$, source-only dissemination requires more than $10\times$ additional steps, increasing from roughly $100$ to over $1000$, whereas peer-Turbo remains much more stable. Similarly, increasing the number of peers from $m=10$ to $m=1000$ raises the steps needed to reach a $\phi=0.8$ quorum by about $8\times$ without peer-Turbo, but only about $2\times$ with peer-Turbo.

\section{Conclusions}
\label{sec:conclude}
We studied RLNC-enabled Peer-Turbo connections for payload dissemination in star-shaped topologies. By allowing destination nodes to exchange RLNC-coded shards, peers can help each other complete decoding without explicit shard-state coordination. We introduced a fluid approximation over degrees of freedom to tractably model system-wide decoding dynamics.

Our results show that Peer-Turbo can substantially reduce dissemination latency, achieving up to an order-of-magnitude faster decoding than source-only dissemination while remaining more robust under reduced source bandwidth.

Future work may extend the degree-of-freedom fluid model to more general network and coded-storage settings, and study its interaction with incentive mechanisms and per-shard pricing~\cite{pricing}.

%%% To balance the columns on the last page, use:
%\enlargethispage{-1.2cm}
%%% Or trigger a column break just before a specific reference number:
%\IEEEtriggeratref{7}

%%%
%%% BIBLIOGRAPHY
%%% (An optional 6th page containing only references is permitted.)
%%%
\bibliographystyle{IEEEtran}
\bibliography{GGBiblio}

\end{document}